\documentstyle[12pt,aasms4]{article}

\def\arcsecpoint{$''\!.$}
\def\deg{$^{\rm o}$}

\received{}
\accepted{}

\lefthead{Crenshaw et al.}
\righthead{Kinematics in NGC 4151}

\begin{document}

\title{A Kinematic Model for the Narrow-Line Region in NGC 4151\altaffilmark{1}}

\author{D.M. Crenshaw\altaffilmark{2,3},
S.B. Kraemer\altaffilmark{2},
J.B. Hutchings\altaffilmark{4},
L.D. Bradley II\altaffilmark{5},
T.R. Gull\altaffilmark{6},\\
M.E. Kaiser\altaffilmark{5},
C.H. Nelson\altaffilmark{7},
J.R. Ruiz\altaffilmark{2},
\& D. Weistrop\altaffilmark{7}
}

\altaffiltext{1}{Based on observations made with the NASA/ESA Hubble Space 
Telescope. STScI is operated by the Association of Universities for Research in 
Astronomy, Inc. under NASA contract NAS5-26555. }

\altaffiltext{2}{Catholic University of America and Laboratory for Astronomy and 
Solar Physics, NASA's Goddard Space Flight Center, Code 681,
Greenbelt, MD  20771}

\altaffiltext{3}{crenshaw@buckeye.gsfc.nasa.gov}

\altaffiltext{4}{Dominion Astrophysical Observatory, National Research
Council of Canada, 5071 W. Saanich Rd., Victoria, B.C. V8X 4M6, Canada}

\altaffiltext{5}{Department of Physics and Astronomy, Johns Hopkins University,
Baltimore, MD 21218}

\altaffiltext{6}{NASA's Goddard Space Flight Center, Laboratory for Astronomy
and Solar Physics, Code 681, Greenbelt, MD 20771}

\altaffiltext{7}{Department of Physics, University of Nevada, Las Vegas,
4505 Maryland Parkway, Las Vegas, NV 89154-4002}

\begin{abstract}
We present a simple kinematic model for the narrow-line region (NLR) of the 
Seyfert 1 galaxy NGC 4151, based on our previous observations of extended 
[O~III] emission with the Space Telescope Imaging Spectrograph (STIS). 
The model is similar to a biconical radial outflow model developed for the 
Seyfert 2 galaxy NGC 1068, except that the bicone axis is tilted much more into 
our line of sight (40\deg\ out of the plane of the sky instead of 5\deg), and 
the maximum space velocities are lower (750 km s$^{-1}$ instead of 1300 km 
s$^{-1}$). We find evidence for radial acceleration of the emission-line knots 
to a distance of 160 pc, followed by deceleration that 
approaches the systemic velocity at a distance of 290 pc (for a distance to 
NGC~4151 of 13.3 Mpc).
Other similarities to the kinematics of NGC 1068 are: 1) there are a number of 
high-velocity clouds that are not decelerated, suggesting that the medium 
responsible for the deceleration is patchy, and 2) the bicone in NGC~4151 is 
at least partially evacuated along its axis.
Together, these two Seyfert galaxies provide strong 
evidence for radial outflow (e.g., due to radiation and/or wind pressure) and 
against gravitational motion or expansion away from the radio jets as the 
principal kinematic component in the NLR.

\end{abstract}


\section{Introduction}

The study of the kinematics of the narrow line region (NLR) in Seyfert galaxies 
is important, because it has the potential to reveal the dynamical forces at 
work on the emission-line gas within about one kpc of the nucleus. In this 
region, the central supermassive black hole presumably dominates the kinematics, 
due to its gravitational influence and/or the radiation, winds, and jets that 
emanate from its immediate surroundings. With the Hubble Space Telescope (HST), 
we have the ability to resolve the emission-line knots within the NLR and with 
the two-dimensional detectors of the Space Telescope Imaging Spectrograph 
(STIS), we can measure their velocities accurately and efficiently.
In this paper, we concentrate on STIS spectra of the apparently brightest 
Seyfert 1 galaxy, NGC 4151, and compare our results to those from a kinematic 
study based on STIS spectra of the brightest Seyfert 2 galaxy, NGC 1068
(Crenshaw \& Kraemer 2000a, hereafter CK2000).

The NLR of NGC 4151 as defined by HST [O~III] images (Evans et al. 1993) is 
biconical in shape with the apex located at the bright nucleus,
and contains numerous bright emission-line knots and filaments.
Evans et al. find a projected half-opening angle of 37.5\deg\ $\pm$ 5\deg\ for 
the bicone and a position angle (PA) of the bicone axis on the sky of 60\deg\ 
$\pm$ 5\deg. 
The extended narrow-line region (ENLR), at distances greater than 6$''$ from the 
nucleus, lies within the extrapolated bicone, but most of the emission from this 
region is concentrated between position angles 30\deg\ and 60\deg\ 
(P\'{e}rez-Fournon \& Wilson 1990). Ground-based spectroscopy has revealed that 
the [O~III] and H~I velocity fields are very similar at these distances (Pedlar 
et al. 1992), indicating that the kinematics in the ENLR are dominated by 
galactic rotation.

Although the velocity field of the NLR in NGC 4151 (at distances less than 6$''$ 
from the nucleus) is difficult to determine from the ground, Schulz (1990) found 
evidence for outflow along the cones, superimposed on galactic rotation. 
With observations primarily from HST's Faint Object Camera (FOC), which has a 
limited ability to 
obtain long-slit spectra, Winge et al. (1997, 1999) claimed that the dominant 
kinematic component is rotation. STIS observations and detailed kinematic models 
are crucial for 
testing these claims. As part of a STIS Key Project carried out by members of 
the Instrument Definition Team, we have accumulated a number of observations of 
NGC~4151. Our previous results on these data are described in Hutchings et al. 
(1998, 1999), Kaiser et al. (2000), Nelson et al. (2000), and Kraemer et al. 
(2000). We describe the observations that are pertinent to our kinematic model 
in the next section.

We take the systemic velocity of NGC~4151 to be 997 km s$^{-1}$, based on H~I 
observations of the outer regions of the host galaxy (Pedlar et al. 1992). For a 
Hubble constant of H$_{0}$ $=$ 75 km s$^{-1}$ Mpc$^{-1}$ and a distance to 
NGC~4151 of 13.3 Mpc, 0\arcsecpoint1 corresponds to 6.4 pc. 

\section{Previous STIS Observations and Results}

Our first STIS observation was a slitless spectrum of the H$\beta$ and 
[O~III]$\lambda\lambda$4959, 5007 emission at a spectral resolving power of 
$\lambda$/$\Delta\lambda$ $\approx$ 8000 (Hutchings et al. 1998). By comparing 
the slitless spectrum with an archival WFPC2 image obtained through a 
narrow-band [O~III] filter, we were able to 
determine the radial velocities and velocity dispersions of about 40 bright 
emission-line knots (with sizes on the order of 0\arcsecpoint2). We could also 
identify about 10 knots in a slitless spectrum of the C~IV region, and found 
that their velocities were consistent with the 
[O~III] measurements. We found that the highest velocities are close to the 
nucleus, and, in general, the emission-line knots are blueshifted to the
southwest and redshifted to the northeast. A rotation curve 
cannot match the observed radial velocities as a function of distance. We 
proposed a simple model consistent with the geometry of Evans et al. (1993) to 
explain the gross kinematics: radial outflow along the surface of a bicone, with 
an opening angle of 80\deg\ and a line of sight that is 10\deg\ outside of the 
cones.

In Kaiser et al. (2000), we presented slitless spectra of several spectral 
regions at a more favorable roll angle than that of the Hutchings et al. (1998) 
observations. Using these data, we were able to determine radial velocities for 
a number of additional emission-line knots. In Figure 1, we compare the radial 
velocities for the 
emission-line knots in common in these two studies.  The average difference in 
radial velocity is only 47 km s$^{-1}$, which indicates the reliability of this 
observational technique. In Kaiser et al. (2000), an attempt to fit the radial 
velocities as a function of distance with a rotating disk model and point-source 
mass was not satisfactory; the transition between blueshifted and redshifted 
velocities is much steeper than observed, and an unreasonably large central mass 
is required: 10$^{10}$M$_{\odot}$ if all clouds are included, and 4 x 
10$^{8}$M$_{\odot}$ if clouds with radial velocities of only 200 km s$^{-1}$ or 
less are included. Thus, we found that outflow is extremely likely, since 
gravitational motions (including infall) cannot account for the high velocities. 
There is no strong correlation between velocity or velocity dispersion 
and the positions of the radio knots, which indicates that jet acceleration is 
not important; wind or radiation pressure are more likely sources of the 
acceleration.

In Hutchings et al. (1999), we used narrow-band WFPC2 images of NGC~4151 to 
isolate faint line emission at high velocities for several ions. Radial 
velocities up to 1400 km s$^{-1}$ were seen in both approach and recession on 
both sides of the nucleus, indicating a separate kinematic component from that 
seen in the bright emission-line knots. We will not attempt to model this 
component in the current paper; additional progress on this component is likely 
to come from planned long-slit observations at high spectral resolution.

In Nelson et al. (2000), we presented long-slit (52$''$ x 0\arcsecpoint1) 
spectra of NGC 4151, obtained at 
low spectral resolving power ($\lambda$/$\Delta\lambda$ $\approx$ 1000) over the 
1150 -- 10,000 \AA\ range at two position angles: 221\deg\ (passing through the 
nucleus) and 70\deg\ (passing 0\arcsecpoint1 south of the nucleus, see Figure 1 
in Nelson et 
al. 2000 for the slit placements superposed on WFPC2 [O~III] images).
Although these data do not have the  spatial coverage and spectral resolution of 
the slitless spectra, the slit allowed us to determine the location and
radial velocities of much weaker emission, including that between the bright 
emission-line knots.
To determine the radial velocities, we used the [O~III] $\lambda$5007 emission 
in 0\arcsecpoint2 bins along the 0\arcsecpoint1-wide slit. We explored two basic 
models in this paper: radial outflow away from the nucleus and 
expansion away from the radio jet axis. We concluded that radial outflow 
provides the best match with the data.

\section{Data Analysis and Measurements}

We rely on our previous observations to develop a detailed kinematic model for 
the NLR of NGC~4151 in this paper.
For our analysis of the slitless spectra, we use the measurements of Kaiser et 
al. (2000), which we have shown are consistent with those of Hutchings et al. 
(1998). For the long-slit spectra, we remeasured the [O~III] $\lambda$5007 
emission along the slit at 
each pixel location (0\arcsecpoint05 intervals), rather than in 0\arcsecpoint2 
bins, to avoid smoothing of features in the radial velocity curves. At some 
locations close to the nucleus, there are two distinct components of 
the [O~III] emission at this spectral resolution. We fit the spectrum at each 
location with a local continuum plus a Gaussian for each clearly identifiable 
peak of [O~III] emission, resulting in one or two kinematic components. We 
determined uncertainties 
in the radial velocities of the components from the errors in the Gaussian 
fits and from different reasonable continuum placements. 

Figures 2 and 3 give the radial velocities, widths (full-width at half-maximum, 
corrected for the instrumental value of 300 km s$^{-1}$), and fluxes as a 
function of 
angular distance from the nucleus for the long-slit spectra at the two position 
angles; our convention is that negative positions represent the southwest 
direction. The radial velocity curves in the top plots are very similar to those 
determined by Nelson et al. (2000), except they show a little more structure due 
to the finer sampling. At distances less than 4$''$, the emission-line knots 
southwest of the nucleus are blueshifted and the knots northeast of the nucleus 
are redshifted. At greater distances, the radial velocities approach the 
rotational velocities seen in the ENLR. High radial velocities 
(greater than $\sim$300 km s$^{-1}$) tend to occur within $\sim$1$''$ of the 
nucleus, although low radial velocities occur in this region as well.
In many cases, the structure in the radial velocity curves can be attributed to 
the varying contributions of emission-line knots and background emission 
to the total [O~III] emission as a 
function of position. The knotty structure can be seen in the bottom plots of 
Figures 2 and 3. The middle plots in Figures 2 and 3 show that there are no 
strong trends in the velocity widths, except that the lower velocity clouds at 
large distances tend to have smaller widths. The velocity widths are due 
to superposition of emission knots with distinct velocities as well as internal 
dispersion within each knot.

In Figure 4, we compare the radial velocities derived from the long-slit spectra 
with those from the slitless spectra; the number of points are limited, since 
there only a few 
bright emission-line knots that lie along the slit locations. Again, the 
measured radial 
velocities are very similar, with no evidence for systematic offsets as a 
function of velocity or dispersion. The average difference in radial velocity is 
only 37 km s$^{-1}$. We conclude that the two observational 
techniques provide reliable radial velocities; slitless spectra at moderate 
resolution are useful for obtaining velocities of all of the bright 
emission-line knots in a single observation, whereas long-slit low-resolution 
spectra are helpful for determining the kinematics of all of the 
emission along a particular direction. 

\section {A Biconical Outflow Model}

Our model for NGC 4151 is based on a previous kinematic model for the NLR in
the Seyfert 2 galaxy NGC 1068 (CK2000). From STIS 
low-resolution long-slit spectra of the [O~III] emission, we found that a simple 
model of biconial radial outflow provides a good match to the radial velocities 
as a function of distance from the hot spot in NGC 1068. In that model, the 
bicone is evacuated along its axis, which is inclined only 5\deg\ out of the 
plane of the sky. The emission-line knots show radial acceleration to a distance 
of $\sim$140 pc from the hot spot, followed by deceleration that approaches the 
systemic velocity at $\sim$310 pc. The deceleration is likely due to collision 
with a patchy and anisotropically distributed ambient medium.

For the biconical model of NGC 4151, we assume that the two cones have 
identical properties (geometry, velocity law, etc.), a filling factor of one 
within the defined geometry, and no absorption of [O~III] photons. The 
parameters that are allowed to vary in our code are the extent of each cone 
along its axis (z$_{max}$), its minimum and maximum half-opening angles 
($\theta$$_{inner}$ and $\theta$$_{outer}$), the inclination 
of its axis out of the plane of the sky (i$_{axis}$), and the velocity law as a 
function of distance from the nucleus. Our code generates a two-dimensional 
velocity map and samples this map with a slit that matches the position, 
orientation, and width of the observational slits. Note that we are only 
attempting to match the observed radial velocities with the model, and not the 
observed line widths or fluxes.

Given the success of our model for NGC~1068, we assumed the same velocity law 
for NGC 4151: constant acceleration to a 
maximum velocity (v$_{max}$) at a turnover radius (r$_t$), followed by a 
constant deceleration to zero velocity at the greatest extent of the cone ($=$ 
z$_{max}$/cos $\theta$$_{outer}$). Then we adjusted the other parameters of the 
model to determine if a reasonable match to the observed radial velocities could 
be obtained. First, we tilted the bicone axis out of the sky until 
the observed form of the model radial velocity curves matched that of the 
observations. For NGC 4151, this occurs when the bicone moves completely out of 
the plane of the sky, so that one cone is blueshifted and the other side is 
redshifted. Then, we varied the other parameters until we obtained a reasonable 
match with the observations. Since the observed radial velocity curves show 
significant scatter, we did not attempt to fine-tune the models; instead we 
settled for an illustrative model that matches the overall trends. 

The parameters for our kinematic model of NGC 4151 are shown in Table 1; values 
for the model of NGC 1068 are given for comparison. The bicone 
opening angles, maximum extent, and turnover radius are very similar for the 
two models. The maximum velocity at the turnover is much lower in NGC~4151, 
which may be due to its lower intrinsic luminosity (see the Discussion). The 
principal difference between the two models is the viewing angle, with the 
bicone axis of NGC~4151 much closer to the line of sight. However, as we note in 
the Discussion, the inclination is not enough to place the line of sight into 
the ionization cone. Given the inclination angle, we find that the {\it 
projected} maximum half-opening angle onto the plane of the sky is 43.5\deg, 
which is close to the 37.5\deg$\pm$5\deg\ determined from the HST images by 
Evans et al. (1993).

Two-dimensional velocity maps for the model of NGC~4151 are given in 
Figure 5, along with the slit positions and widths that correspond to our 
long-slit 
observations. For demonstration, we show only the front and back sides 
of the outermost surface (at $\theta$$_{outer}$). For comparison, we show the 
same surfaces in Figure 6 for our model of NGC~1068 (not shown in CK2000 due to 
space limitations). In both figures, the models are aligned so that the 
projected bicone axis on the sky is vertical, and northeast is at the top. The 
inclination of the bicone axis is NGC 4151 is such that the entire northeast 
cone is redshifted, and the entire southwest cone is blueshifted.  By contrast, 
the axis for NGC 1068 is inclined by only 5\deg\, so that blueshifts and 
redshifts of similar magnitude occur in both cones (see CK2000 
for more details). The effects of our acceleration plus deceleration velocity 
laws, modified by projection of the velocity vectors from the bicone, 
are visible in these images.
One interesting feature of the NGC~4151 model is that the 
maximum radial velocities occur very close to the nucleus on the plane of the 
sky, regardless of slit position.

\section{Comparison of the Observations and Models}

First, we compare the models with the long-slit observations.
Figure 7 shows the envelopes of radial velocities from the model of NGC 
4151, extracted from the two slit positions, compared to the observed radial 
velocities (relative to systemic). 
There are four distinct components of the envelope in each plot, corresponding 
to the four quadrants of the bicone intersected by the slit.
The width of the envelope is determined by the range in half-opening angle, and 
the relative amplitudes are determined by the inclination of the bicone axis.
Although the match is not perfect, the overall trends are reproduced.
The major discrepancy is that this model does not explain the blueshifted 
velocities in the $-$150 to $-$300 km s$^{-1}$ range (relative to systemic) in 
the southwest cone. This discrepancy could be eliminated by filling in the 
southwest cone with emitting material, which would fill in the gap in the plots, 
but would also fill in regions that are not occupied by observed points.
Thus, the bicone for the most part must be evacuated along its axis .
Our  model also does not explain the emission knots with very 
high velocities ($>$ 700 km s$^{-1}$ relative to systemic), which is also the 
case for NGC~1068. As noted in CK2000, this 
indicates that some emission-line knots do not experience a significant 
deceleration. 

Next, we compare the models with the slitless observations. In the top plot of 
Figure 8, we show the slitless radial velocities as a function of position, 
relative to the nucleus at the origin. The magnitude of the radial velocity is 
represented by the size of the symbol, with ``+'' representing blueshift and 
``x'' representing redshift, relative to the systemic velocity. The plot is 
oriented such that the bicone axis, at PA = 60\deg, is vertical. The trends 
discussed earlier can be seen in this plot (for a color-coded velocity 
map, see Hutchings et al. 1998).

The bottom plot in Figure 8 shows the observed radial velocities minus the model 
values. At each observed point, we have two model values to choose from:
the average radial velocity for the front side and the average value for the 
back side; we adopt the value closest to the observed radial velocity and 
subtract. Although 
our kinematic model was calculated to fit the long-slit data, the small residual 
velocities in this plot show that it provides a reasonable fit to the slitless 
data as well. There is a mixture of positive and negative 
residuals in each cone, indicating the overall agreement between the model and 
observations. The two largest residuals in this plot
identify the ``rogue'' blueshifted knots that apparently do not 
experience any deceleration, which  correspond to knots 23 and 26 in 
Hutchings et al. (1998) and Kaiser et al. (2000).

In Figure 9 we plot the observed radial velocities vs. the model values at each 
point to show the magnitude of the residuals. Except for a few points, notably 
the two highly blueshifted knots, the residuals are relatively small and evenly 
distributed 
around the unity line. We conclude that our kinematic model provides a good 
match to the general trends seen in all of our slitless and long-slit data sets.

\section{Discussion}

The details of our kinematic model of NGC~4151 were inspired by STIS 
observations and a kinematic model of the NLR in NGC~1068. Since the outflow 
axis is nearly in the plane of the sky in NGC~1068, the trend of increasing 
velocity with distance to a turnover radius, followed by decreasing velocity, 
can be easily seen in this object, and the lack of emission along the cone axis 
in NGC~1068 is also clear (CK2000). The detailed kinematics of the NLR in NGC 
4151 have been more difficult to interpret, since it turns out that the cone 
axis is significantly out of the plane of the sky, and velocities from different 
radial distances are projected along the same line of sight (see Figure 5). 
However, by adopting the basic geometry and velocity law used 
for NGC~1068, we are able to construct a kinematic model of the NLR in NGC 4151 
that matches the observed trends in radial velocity as a function of position. 
The only major changes were to reduce the maximum radial velocity from 
1300 to 750 km s$^{-1}$, and to tilt the bicone axis from 5\deg\ to 40\deg\ out 
of the plane of the sky. Thus, we have a basic model that can explain the NLR 
kinematics in both Seyfert galaxies.

\subsection{Comparison with Other Models}

Winge et al. (1997, 1999) claim that the dominant kinematic component in the NLR 
of NGC 4151 is rotation. However, there are several problems with this model. 
First, as we discussed earlier, the observed transition between blueshifted and 
redshifted radial velocities is much shallower than the predicted values from a 
Keplerian disk with a point-source mass. Thus, in addition to a point-source 
mass of 5 x 10$^{7}$ M$\odot$, these authors must include an extended mass of 
order 10$^{9}$ M$_{\odot}$ in the inner 0\arcsecpoint1, to match the shallow 
transition between blueshifted and redshifted velocities. To date, there is no 
confirmatory evidence for such an extended mass. Second, Winge et al. (1999) are 
only able to fit their low-velocity 
component (within 300 km s$^{-1}$ of systemic) with this model; the higher 
velocities require another explanation. Finally, we note that a rotation model 
cannot explain the kinematics of the NLR in NGC 1068, which shows blueshifts and 
redshifts of similar magnitude on both sides of the nucleus.

To explore the uniqueness of our model for NGC 4151, we have experimented with a 
number of different velocity laws and inclination angles, but with the same 
biconical geometry. We summarize our findings as follows.
1) Simple radial outflow at constant velocity or acceleration cannot 
explain the decreasing velocities at distances $>$ 1$''$. 
2)~Gravitational infall cannot produce the small radial 
velocities seen within 1$''$ of the nucleus. Also, as we pointed out in Kaiser 
et al. (2000), gravitational infall models require unreasonably high masses 
(on the order of 10$^{10}$ M$\odot$) to produce the high velocities that are 
seen.
3) Axon et al. (1998) propose a model for NGC 1068 in which the gas expands away 
from the radio jet (which is nearly coincident with the ionization cone axis). 
The simplest version of this model is one in which the velocity vectors are 
perpendicular to the bicone axis. As demonstrated by Nelson et al. (2000), this 
model cannot work for NGC~4151. As the bicone axis is 
tilted out of the plane of the sky, the velocity vectors in each of the two 
cones decrease in magnitude by the same amount, and there is no inclination 
angle that will yield the observed blueshifts on one side of the nucleus and 
redshifts on the other.
4) Nelson et al. (2000) show that a reasonably good match to the STIS 
observations can be obtained by assuming radial outflow, with velocity 
decreasing with distance as r$^{-1/2}$. We verify that this velocity law 
provides an acceptable solution for NGC~4151, except that there is some 
indication of acceleration of the clouds away from the nucleus at small 
distances (see Figure 7). We note that this velocity law 
does not work for NGC~1068, which shows clear acceleration of the clouds to an 
angular distance of $\sim$1.7$''$ (CK2000). Thus, our proposed model provides 
the best explanation at present for both Seyfert galaxies.

\subsection {Geometry of the Bicone and Host Galaxy}

The host galaxy of NGC 4151 is nearly face-on (inclination from our line of 
sight $=$ 20\deg,  PA of the major axis $=$ 22\deg, see Pedlar et al. 1992).
Given the inclination and PA of our kinematic model of the NLR, the bicone axis 
is inclined by $\sim$40\deg\ with respect to the host galaxy, and one side of 
each cone (corresponding to a low radial velocity quadrant in Figure 7) lies 
close to the galactic plane (similar to the case for NGC~1068, see CK2000). 
Thus, our proposed geometry allows the gas in the galactic 
disk to be ionized by the continuum source and is compatible with the 
idea that this gas is the ENLR. The general agreement between ground-based 
[O~III] and H~I 
radial velocities at a distance of 10 -- 20$''$ from the nucleus provides strong 
evidence for this idea (Pedlar et al. 1992); in particular, these observations 
indicate normal galactic rotation, with typical radial velocities of $-$70 km 
s$^{-1}$ in the southwest and $+$70 km s$^{-1}$ in the northeast.
Our slitless observations are consistent with these values; they indicate an 
average radial velocity of $-$50 $\pm$ 15 km s$^{-1}$ relative to systemic 
for the [O~III] emission 10 -- 20$''$ southwest of the nucleus (after correcting 
for an offset of $-$40 km s$^{-1}$ for the [O~III] emission at the nucleus, 
relative to the systemic, see 
Kaiser et al. 2000). Thus, our observations confirm that the NLR and ENLR are 
two physically distinct regions in NGC~4151; from the kinematics, it appears 
that the transition from one region to the other occurs in the region 6$''$ to 
10$''$ from the nucleus.

The standard unified model for Seyfert galaxies posits a thick molecular torus 
that produces ionization cones with sharp edges (Antonucci 1993). In this 
scheme, our line of sight to a Seyfert 1 galaxy should be inside the ionization 
bicone. However,the sum of the inclination and maximum half-opening angles from 
our model of NGC~4151 is 76\deg\, which places our line of sight 14\deg\ outside 
of the bicone. This discrepancy was pointed out Evans et al. (1993), who 
arrive at a viewing angle that is $\sim$30\deg\ outside of the bicone,
based on the observed morphologies of the NLR and ENLR.
Thus, NGC~4151 does not appear to fit the standard unified model. However,
if the obscuring torus is not optically thick at all polar angles (cf., Evans et 
al. 1993; Konigl and Kartje 1994), then the ionization cones would have 
``fuzzy'' edges, and NGC~4151 would offer no problems for this type of model.

\subsection{Implications for Dynamical Models}

Based on our observations and kinematic models, we find strong evidence for 
radial outflow in the NLRs of NGC 4151 and NGC 1068. The velocity flow is 
well-organized, suggesting that the emission-line clouds originate close to the 
nucleus and are forced outward from there. A linear increase in velocity with 
distance to a turnover radius provides a reasonable fit to the observations. 
There are two plausible interpretations of this trend. 
One is that the clouds in these two NLRs originated from an explosive outburst 
that occurred $\sim$10$^{5}$ years ago (the crossing time to a distance of 
$\sim$150 pc), which would naturally lead to a linear increase of velocity with 
distance. If this model is generally correct, then the NLRs in some 
Seyfert galaxies should be evacuated near the nucleus; so far there is no 
evidence for this in HST [O III] images (cf., Schmitt \& Kinney 1996).

Another interpretation assumes a steady-state model, in which clouds 
are continuously ejected from the nucleus and experience approximately constant 
acceleration out to the turnover radius. Under this interpretation, the most 
likely sources of acceleration would be radiation and/or wind pressure.
Radiative driving of resonance lines is an attractive means of accelerating the 
clouds, since a strong radiation field is obviously present. However, there is 
also evidence for outflowing winds in the 
form of highly ionized gas seen in absorption in the UV (Crenshaw et al. 1999) 
and X-rays (Reynolds 1997; George et al. 1998).
Since the turnover radius is similar for the two Seyfert galaxies, the lower 
maximum velocity in NGC~4151 indicates a smaller acceleration. If the 
acceleration mechanism is radiation pressure, the lower maximum velocity in 
NGC~4151 may be due to a much lower luminosity, which is $\sim$10\% of the 
intrinsic luminosity of NGC~1068 (Pier et al. 1994).
In any event, future dynamical models will have to explain the roughly constant 
acceleration of the NLR clouds to the turnover point.
 
It is unlikely that the radio jets are responsible for the radial acceleration 
of the emission line knots, since there is no apparent correlation of their 
radial velocities with radio knot location in either NGC~4151 (Kaiser et al. 
2000) or NGC~1068 (CK2000). However, the radio jets may be responsible for 
evacuating the bicones along their axes, since the emission-line knots appear to 
be avoiding the radio knots in the inner NLRs of these two Seyferts (Kaiser et 
al. 2000; Capetti et al. 1997).

An unexpected result of this study and that of CK2000 is that the NLR clouds in 
NGC 4151 and NGC 1068 undergo deceleration at a distance of $\sim$150 pc from 
the nucleus. The deceleration cannot be due to gravity because an unreasonably 
high mass ($\sim$10$^{10}$ M$\odot$) is required at the distances and velocities 
of the turnover point. Following CK2000, the simplest explanation for the 
deceleration is that the clouds experience a drag force due to a surrounding 
ambient medium extending outward from the turnover point. In this 
interpretation, the emission knots with unusually high velocities have not 
experienced drag forces because they have passed through holes in the 
surrounding medium. Additional evidence for collision of the NLR clouds with an 
ambient medium comes from a study of the physical conditions in the NLR of NGC 
1068 (Kraemer \& Crenshaw 2000). Near the location of the velocity turnover in 
the blueshifted region of the northeast cone, we find evidence for an additional 
source of EUV ionizing radiation that is likely to be generated from the shock 
front 
as the clouds encounter the ambient medium. However, we note that these effects 
have not (as yet) been detected at other turnover points in NGC 1068 or in NGC 
4151. Another clue to the ambient medium can be found from observations of the 
extended continuum in NGC 1068 (Crenshaw \& Kraemer 2000b), which show very 
enhanced regions of electron-scattered continua. As pointed out by Kraemer \& 
Crenshaw (2000), these are likely to 
be regions of highly-ionized, low-density (ionization parameter $\approx$ 1, 
electron density $\approx$ 
100 cm$^{-3}$) gas which must nearly fill the bicone at these distances, and are 
therefore likely candidates for the decelerating media. Unfortunately, the 
glare from the nucleus of NGC~4151 prevents us from searching for a scattered 
continuum component in its NLR.

\section{Conclusions}

The principal advantage of a kinematic model that invokes radial outflow in the 
NLR is that it can explain the observed radial velocities in both a Seyfert 1 
galaxy (NGC 4151) and a Seyfert 2 galaxy (NGC 1068). Assuming that radial 
outflow is common in the NLR of all Seyfert galaxies, it is interesting (but not 
surprising) that the kinematic signature of circumnuclear gas within $\sim$500 
pc of an active nucleus is probably different from that surrounding ``inactive'' 
supermassive black holes in nearby galaxies, in which gravitational motions 
dominate (e.g., M84, Bower et al. 1998). This may indicate that once a 
supermassive black hole ``turns on'' (e.g., due to an onset of fueling), 
radiation and/or wind pressure from the active nucleus will dominate the 
kinematics of the NLR.

It is intriguing that very similar geometries and velocity laws can explain the 
observed radial velocities in the two Seyfert galaxies that we have studied. 
This may indicate a common dynamical situation in all Seyferts, or it may just 
be a coincidence. New STIS observations of more Seyfert galaxies will be 
helpful in testing the general applicability of these kinematic models and 
providing the refinements needed for detailed dynamical models.

\acknowledgments
This work was supported by NASA Guaranteed Time Observer funding to the STIS 
Science Team under NASA grant NAG 5-4103.

\clearpage

\figcaption[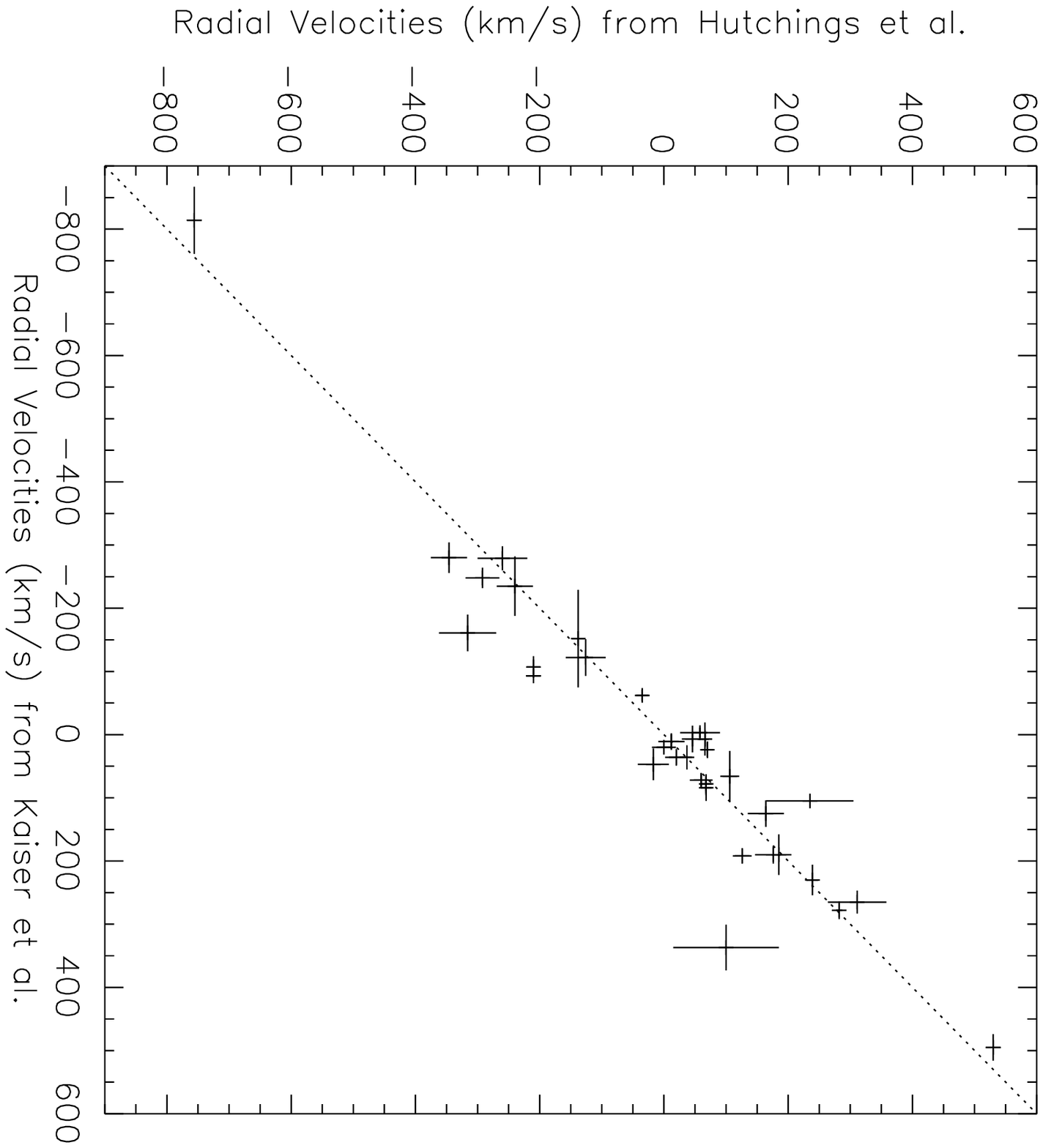]{Comparison of radial velocities from STIS slitless 
spectra 
of the [O~III] emission in NGC 4151, obtained at two different roll angles by 
Hutchings et al. (1998) and Kaiser et al. (2000).}

\figcaption[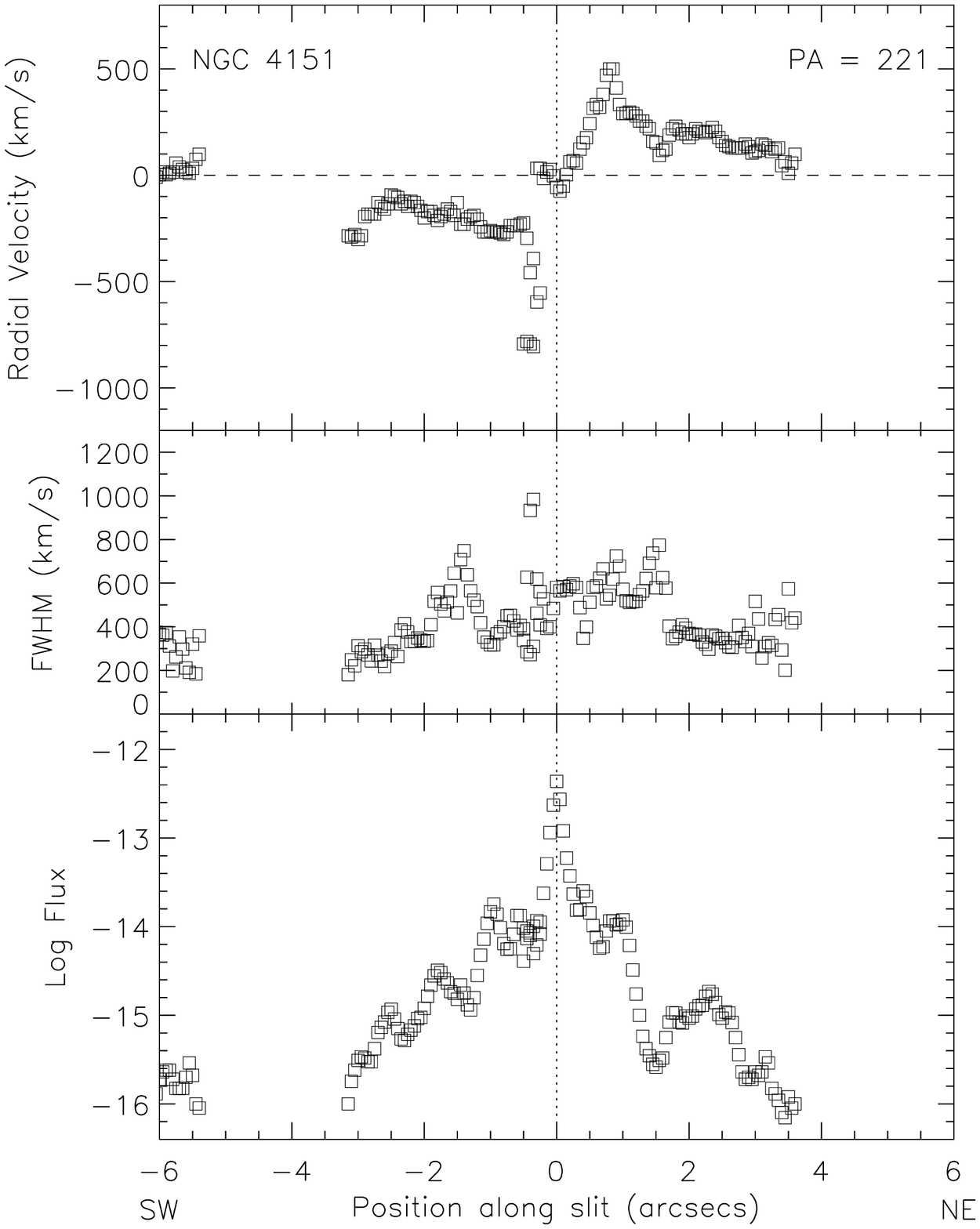]{Measurements of the [O~III] $\lambda$5007 emission 
from 
STIS long-slit low-dispersion spectra, along the position angle 221\deg\ (see 
Nelson et al. 2000). Radial velocities are given relative to the systemic 
velocity from H~I observations (997 km s$^{-1}$). Fluxes are in units of ergs 
s$^{-1}$ cm$^{-2}$ for each 0\arcsecpoint05 x 0\arcsecpoint2 bin.}

\figcaption[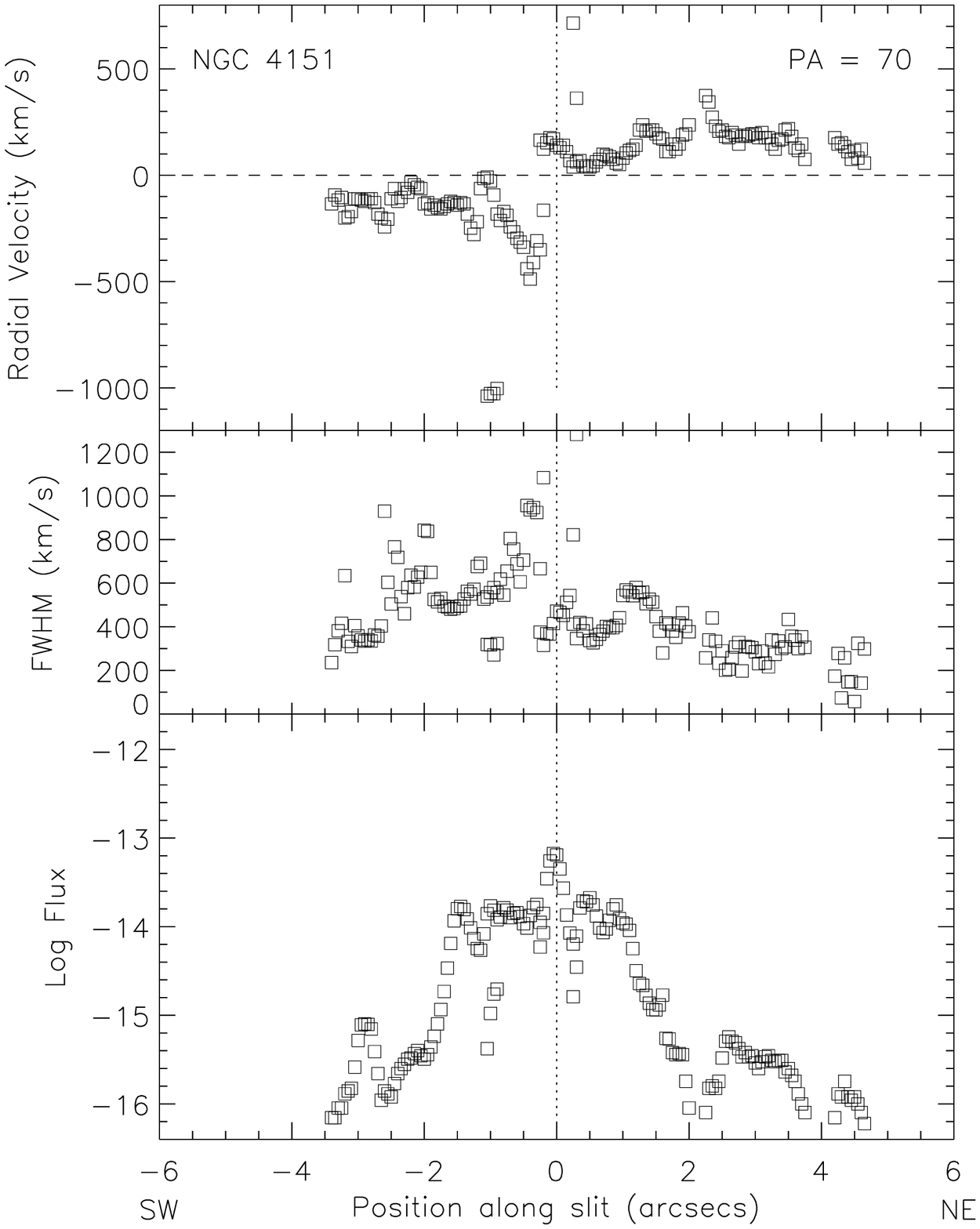]{Same as for Figure 2, except for a slit at 
position angle 
70\deg, offset 0\arcsecpoint1 south of the nucleus.}

\figcaption[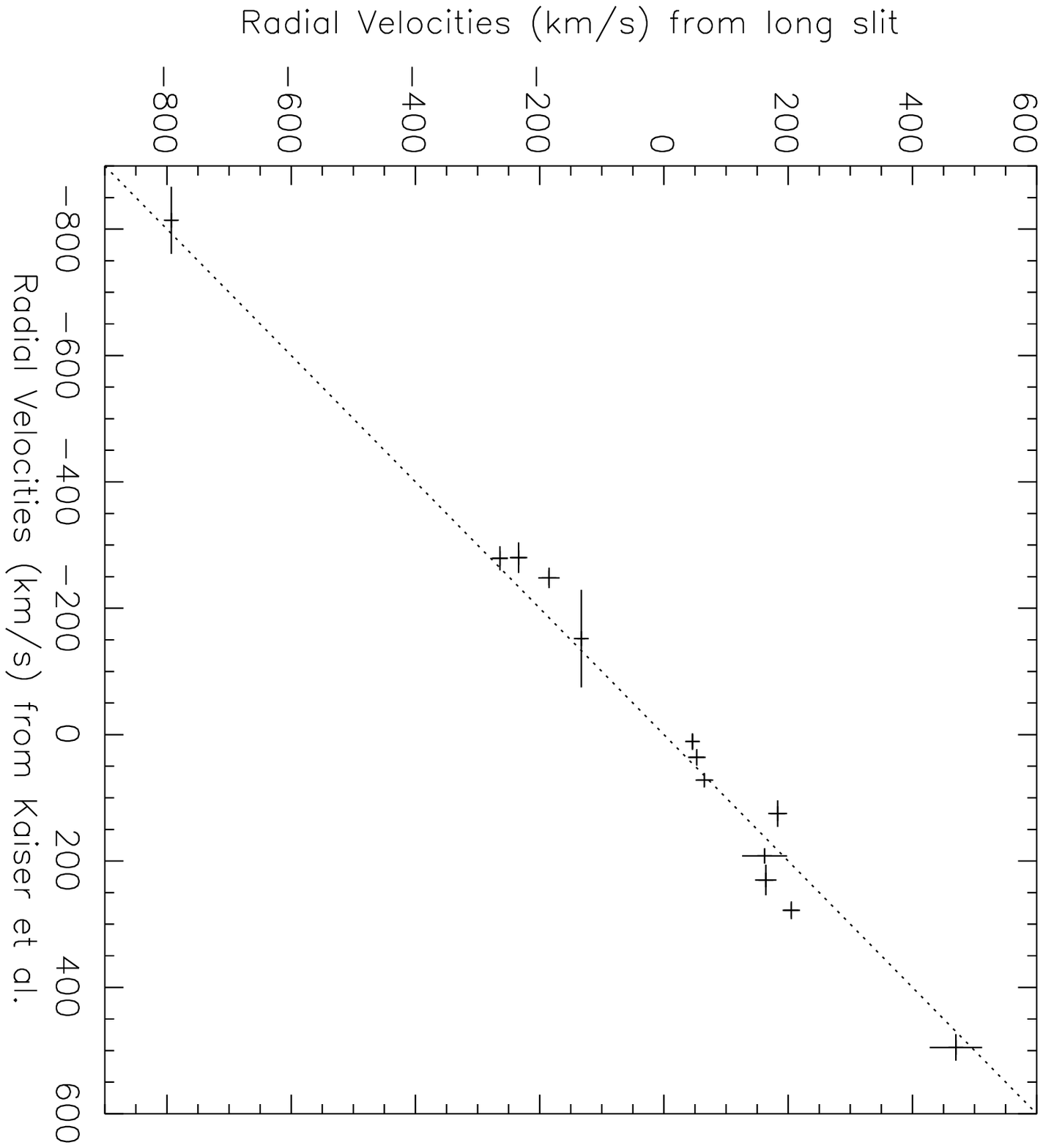]{Comparison of radial velocities from STIS 
long-slit 
observations of the [O~III] $\lambda$5007 emission with those obtained from 
bright emission-line knots in the slitless spectra of Kaiser et al.( 2000) that 
lie along the two slit positions.}

\figcaption[crenshaw.fig5.ps]{Two-dimensional radial velocity maps for the outer 
surface 
of the biconical outflow model of NGC 4151, as they would appear on the sky: 
front side (left image) and back side (right image). The projected kinematic 
axis is vertical (corresponding to a position angle of 60\deg\ in the 
observations, NE at the top), and the relative positions of the 
0\arcsecpoint1-wide slits are 
shown in the near-side image. Deep purple, green, and dark red represent $-$700, 
zero, and $+$700 km sec$^{-1}$, respectively.}

\figcaption[crenshaw.fig6.ps]{Two-dimensional radial velocity maps for the outer 
surface 
of the biconical outflow model of NGC 1068, as they would appear on the sky: 
front side (left image) and back side (right image). The projected kinematic 
axis is vertical (corresponding to a position angle of 15\deg\ in the 
observations, NE at the top), and the relative position of the 
0\arcsecpoint1-wide slit is 
shown in the near-side image. Deep purple, green, and dark red represent $-$900, 
zero, and $+$900 km sec$^{-1}$, respectively (see CK2000 for a comparison with 
the observations).}

\figcaption[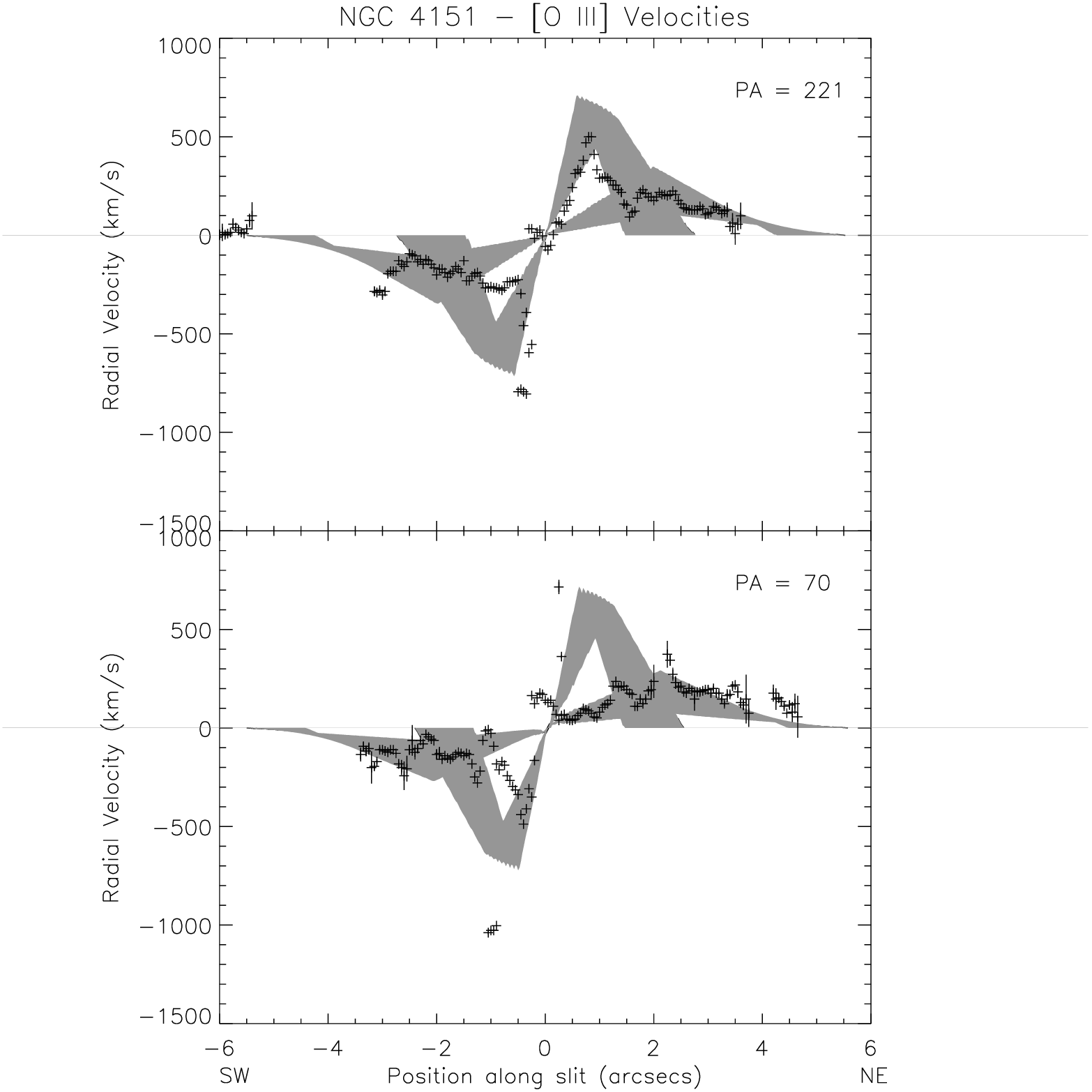]{Comparison of model and observed radial velocities 
from the 
two long-slit positions. Errors in the observed radial velocities are smaller 
than the symbols, except as noted. The shaded regions are the envelopes of 
predicted radial velocities extracted from the slit positions in Figure 5.}

\figcaption[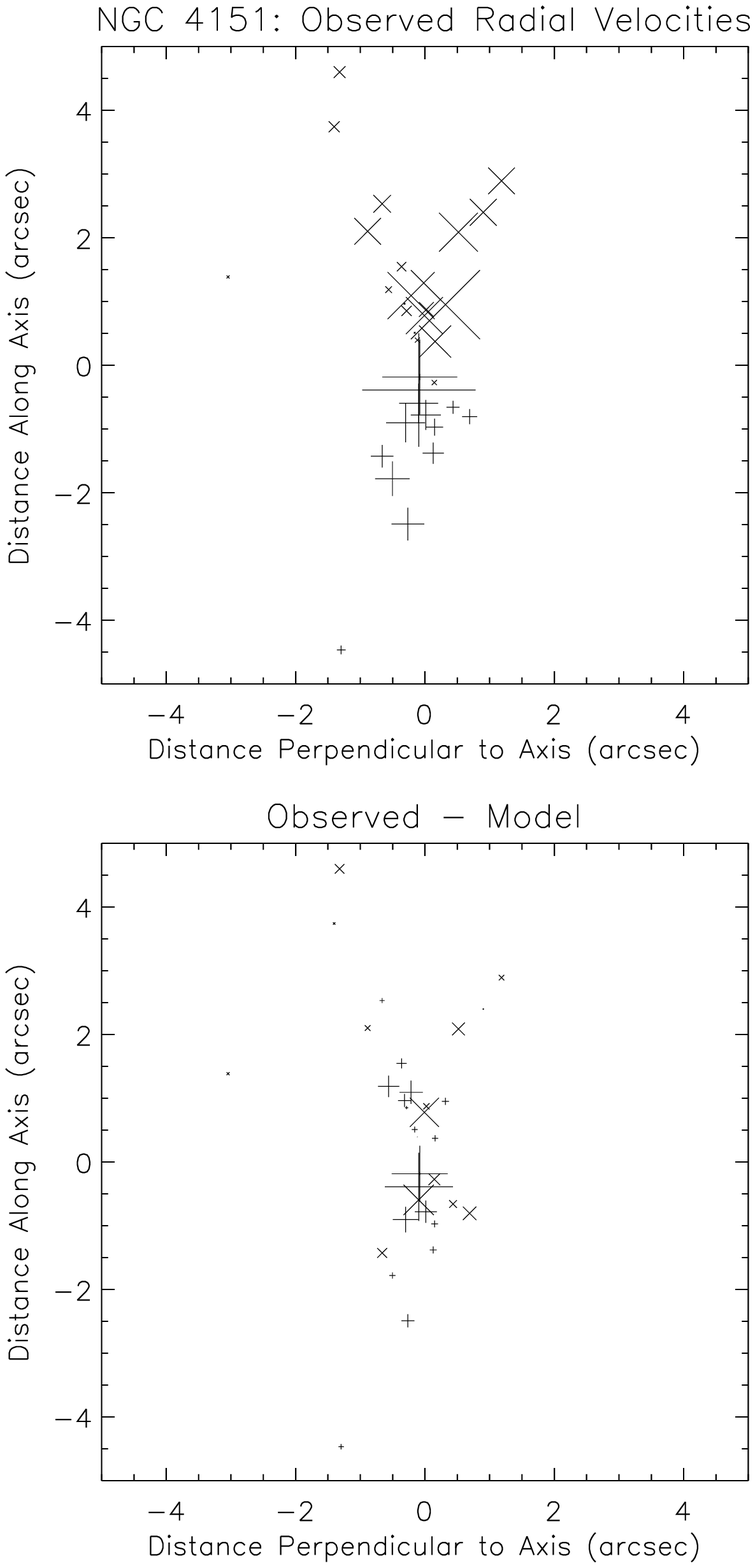]{Two dimensional distribution of radial velocities 
from the 
slitless spectra, with the vertical direction corresponding to the kinematic 
axis (at PA = 60\deg). The size of the symbol represents the magnitude of the 
radial velocity: ``+'' represents blueshifts and ``x'' represents redshifts, 
relative to the systemic velocity. The upper plot shows the observed values, and 
the lower plot shows the residuals (observed - model).}

\figcaption[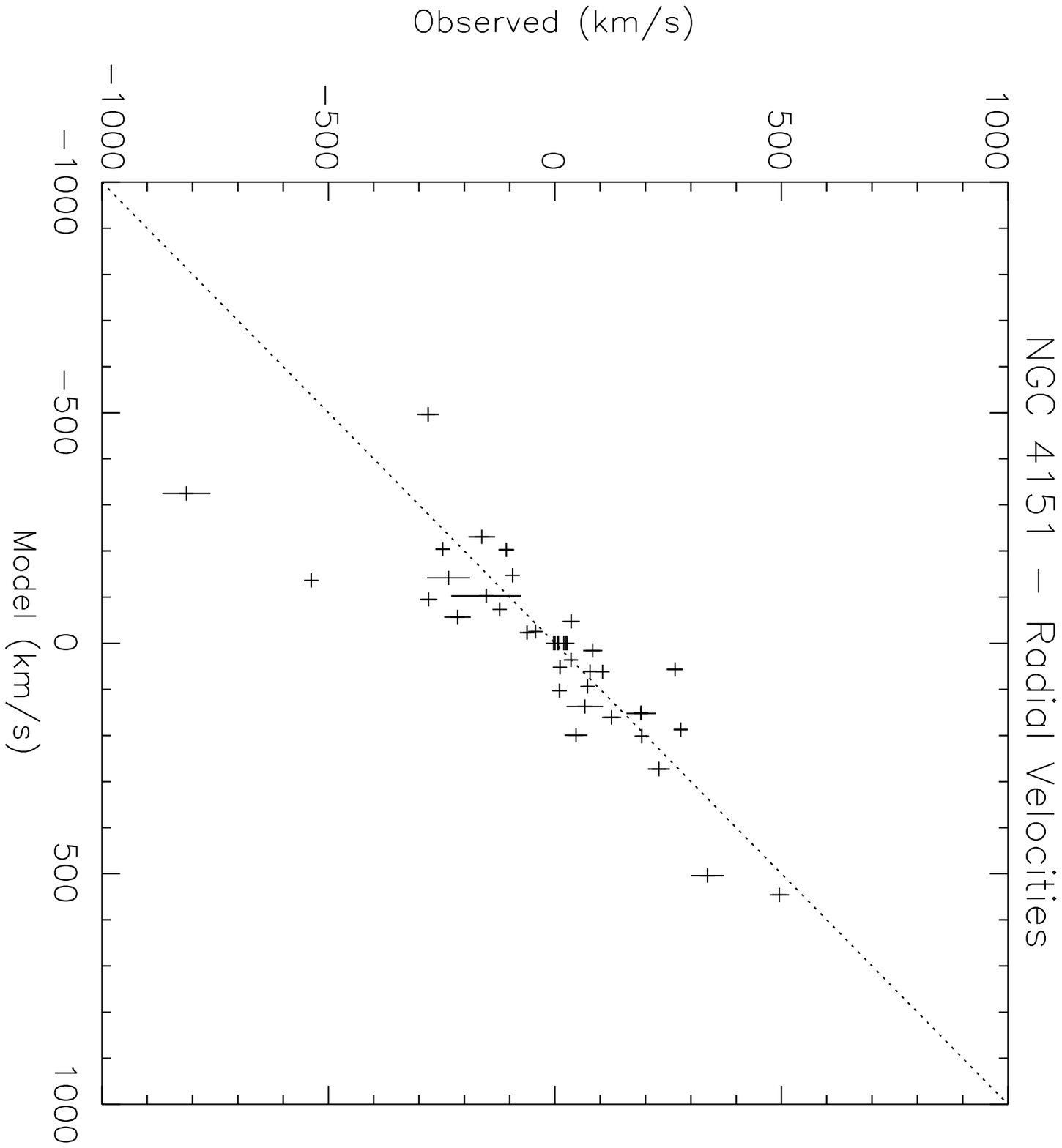]{Comparison of model and observed radial velocities 
from the 
slitless observations (Kaiser et al. 2000).}

\clearpage
\begin{deluxetable}{lll}
\tablecolumns{3}
\footnotesize
\tablecaption{Kinematic Model of NGC 4151 (and NGC 1068)\label{tbl-1}}
\tablewidth{0pt}
\tablehead{
\colhead{Parameter} & \colhead{NGC~4151} & \colhead{NGC~1068}
}
\startdata
z$_{max}$         &288 pc                  &306 pc \\
$\theta$$_{inner}$  &20\deg\               &26\deg\ \\
$\theta$$_{outer}$  &36\deg\               &40\deg\  \\
i$_{axis}$        &40\deg\ (SW is closer)  &5\deg\ (NE is closer)  \\
v$_{max}$         &750 km s$^{-1}$         &1300 km s$^{-1}$\\
r$_t$             &162 pc                  &137 pc \\
\enddata
\end{deluxetable}


\pagestyle{empty}

\clearpage
\vskip3.0in
\begin{figure}
\plotone{crenshaw.fig1.ps}
\\Fig.~1.
\end{figure}

\clearpage
\vskip3.0in
\begin{figure}
\plotone{crenshaw.fig2.ps}
\\Fig.~2.
\end{figure}

\clearpage
\vskip3.0in
\begin{figure}
\plotone{crenshaw.fig3.ps}
\\Fig.~3.
\end{figure}

\clearpage
\vskip3.0in
\begin{figure}
\plotone{crenshaw.fig4.ps}
\\Fig.~4.
\end{figure}

\clearpage
\vskip3.0in
\begin{figure}
\plottwo{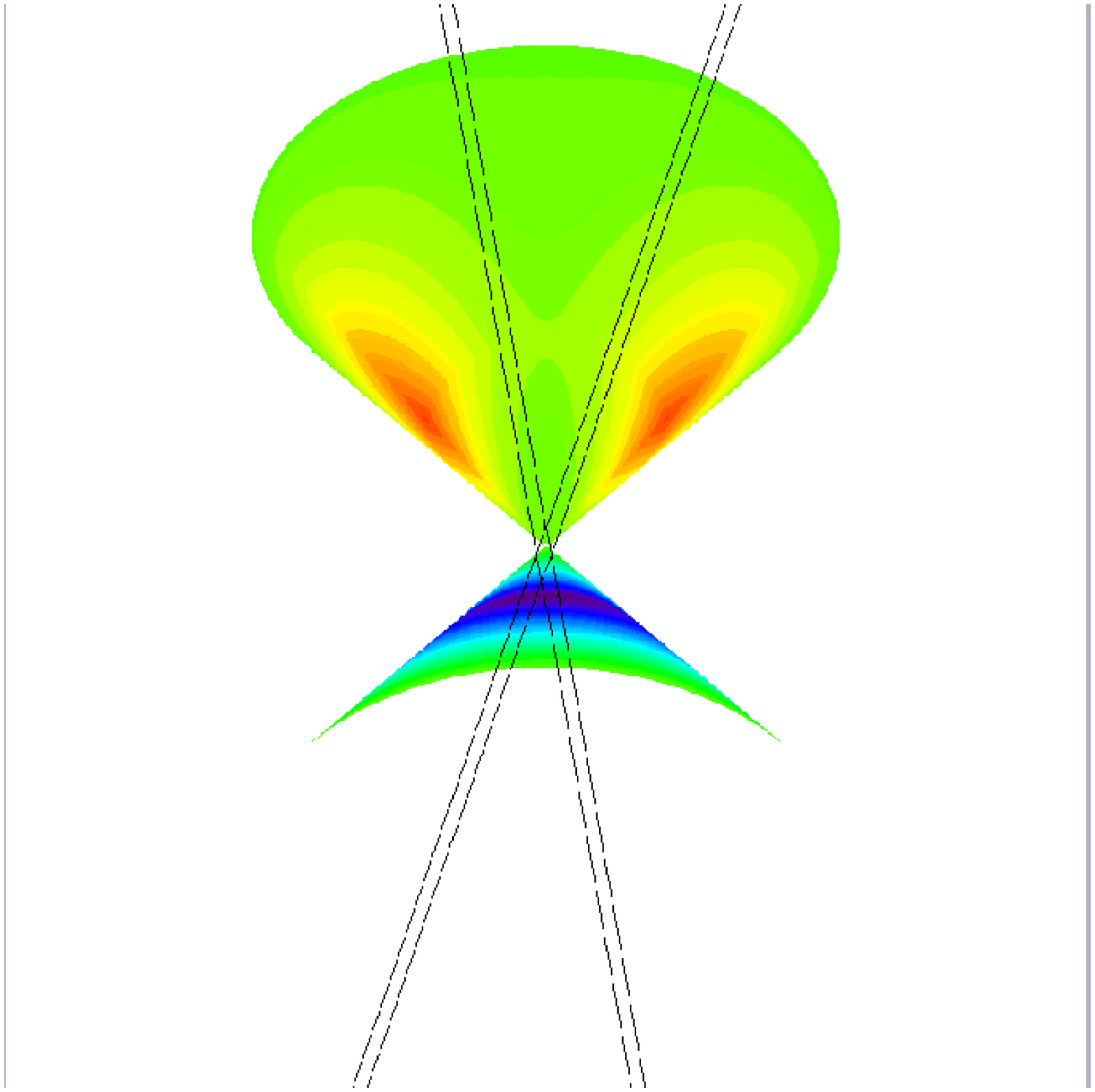}{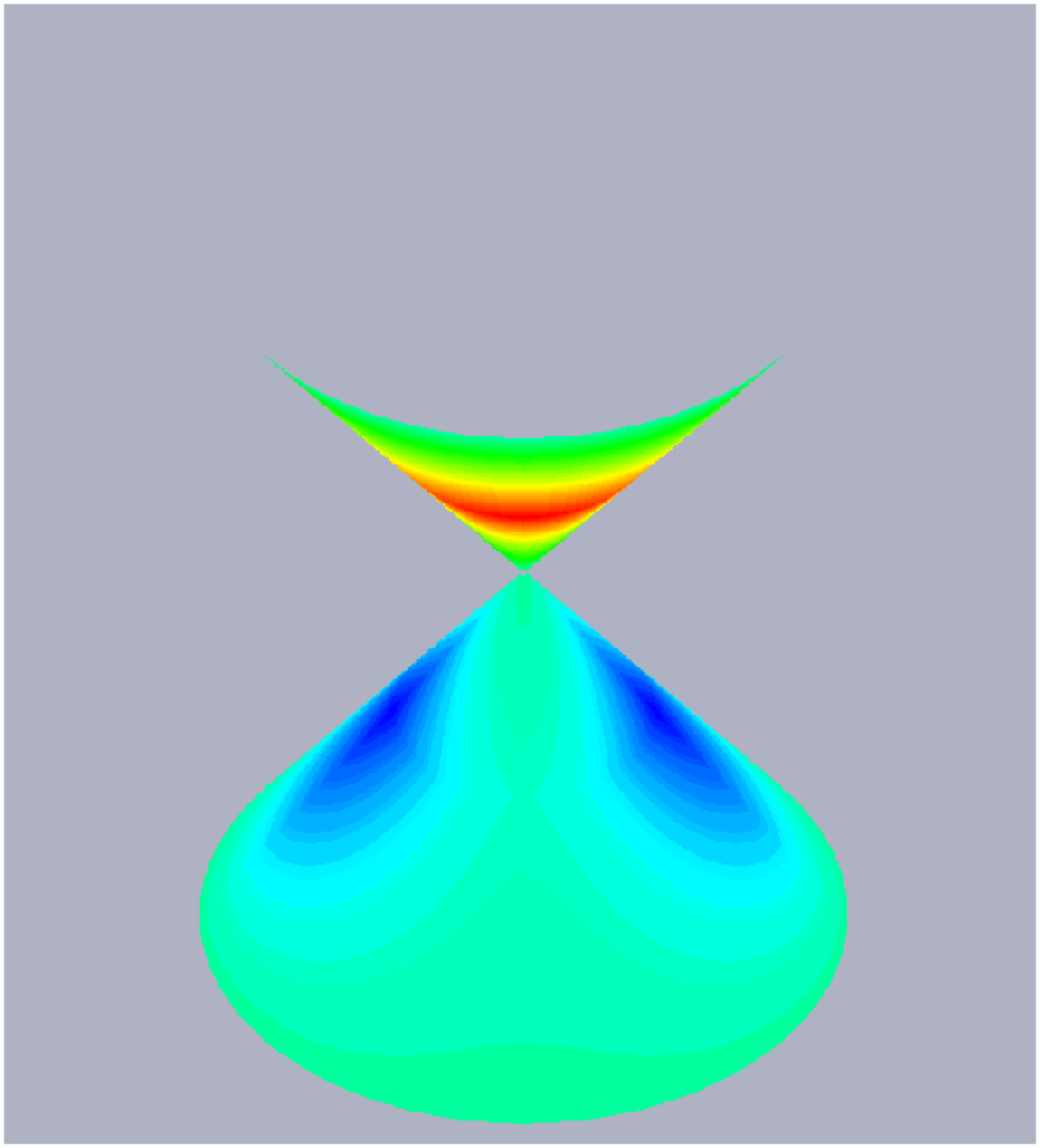}
\\Fig.~5.
\end{figure}

\clearpage
\vskip3.0in
\begin{figure}
\plottwo{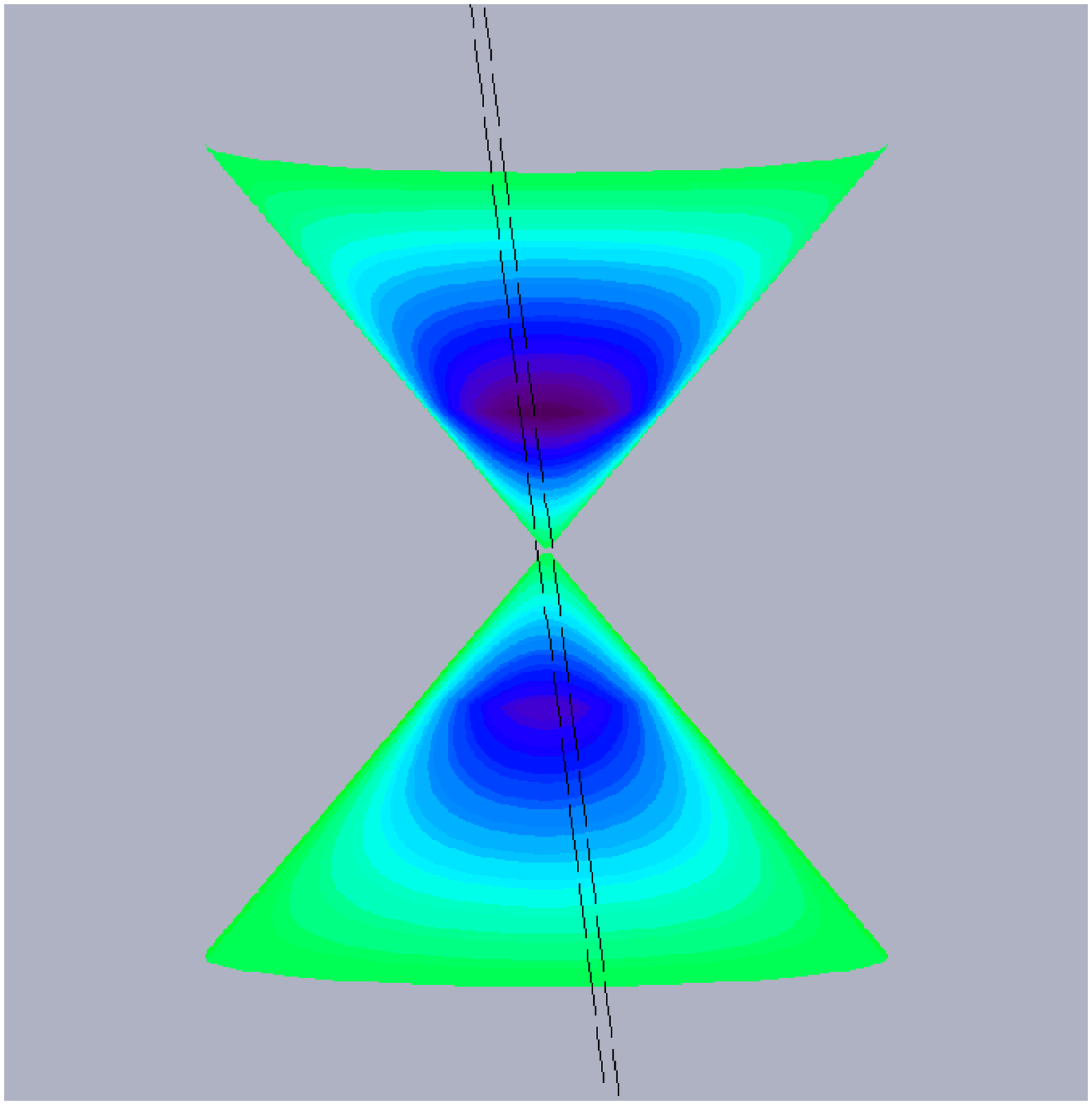}{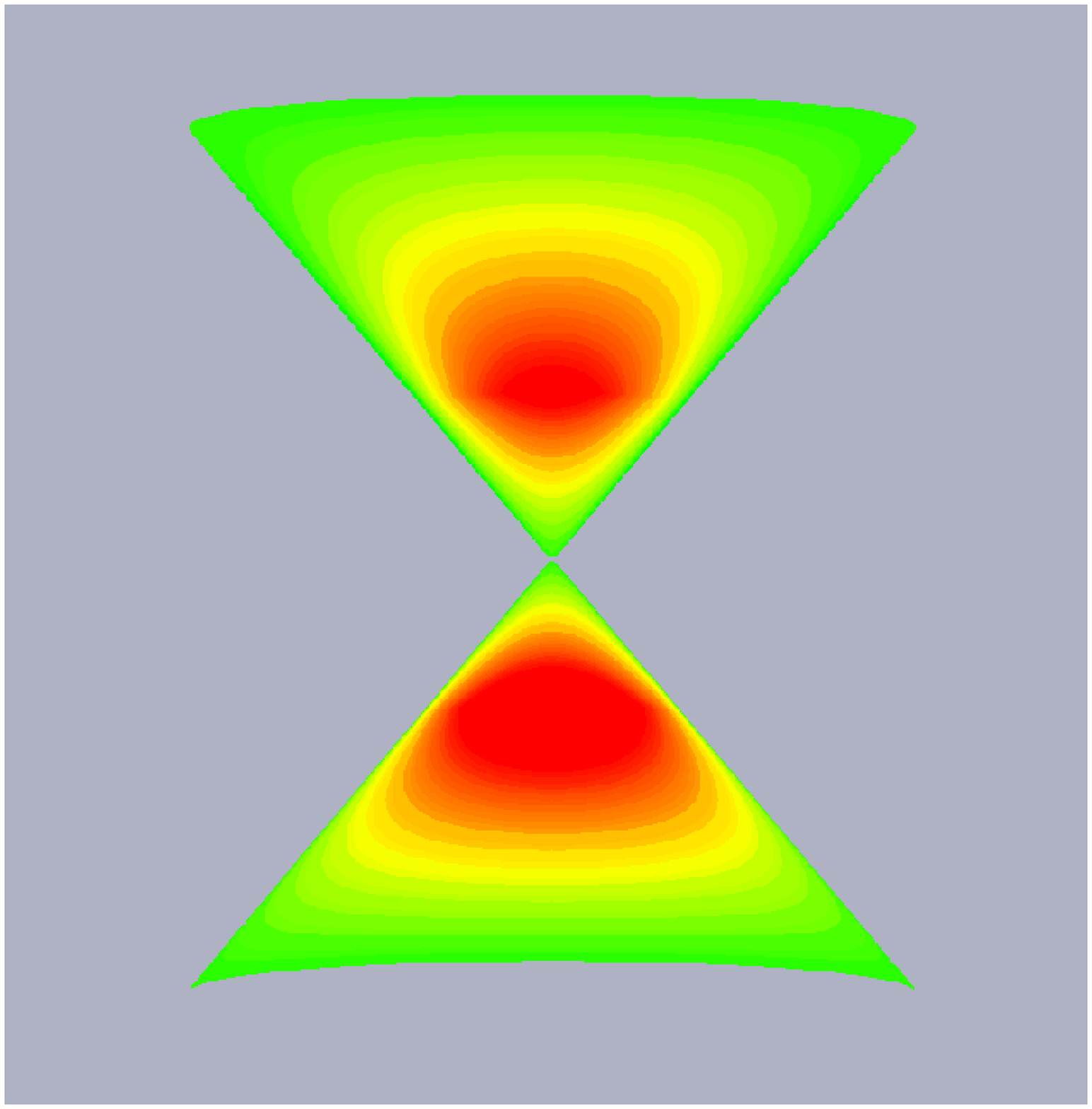}
\\Fig.~6.
\end{figure}

\clearpage
\vskip3.0in
\begin{figure}
\plotone{crenshaw.fig7.ps}
\\Fig.~7.
\end{figure}

\clearpage
\vskip3.0in
\begin{figure}
\plotone{crenshaw.fig8.ps}
\\Fig.~8.
\end{figure}

\clearpage
\vskip3.0in
\begin{figure}
\plotone{crenshaw.fig9.ps}
\\Fig.~9.
\end{figure}

\end{document}